\documentclass[11pt,a4paper]{article}
\pdfoutput=1

\usepackage[utf8]{inputenc}

\usepackage{jheppub}

\usepackage{graphicx,amssymb,amsmath,mathtools,bm,bbm}
\usepackage[labelformat=simple]{subcaption}
\usepackage{autobreak}
\usepackage{hepunits}
\usepackage{physics}
\usepackage{epsfig}

\usepackage{hyperref}
\usepackage{url}
\hypersetup{pdfnewwindow=true}
\usepackage[utf8]{inputenc}

\def\be{\begin{equation}}
\def\ee{\end{equation}}

\allowdisplaybreaks

\preprint{
  \begin{flushright}
    P3H-20-050 \\
    TTK-20-30
  \end{flushright}}

\title{Exact quark-mass dependence of the Higgs-photon form factor at
  three loops in QCD}

\author{Marco Niggetiedt}
\emailAdd{marco.niggetiedt@rwth-aachen.de}
\affiliation{Institut f\"{u}r Theoretische Teilchenphysik und
  Kosmologie,
  RWTH Aachen University, \\
  D-52056 Aachen, Germany}

\abstract{We follow up on our discussion of the exact quark-mass dependence of the Higgs-gluon form factor at three loops in QCD \cite{Czakon:2020vql} and turn our attention to the closely related Higgs-photon form factor. Similarly to our previous work, we intend to examine the form factor for the decay of a Higgs-boson with variable mass into two photons at the three-loop level in QCD. The set of master integrals is known numerically due to prior work on the Higgs-gluon form factor and is exploited to obtain expansions around the threshold as well as in the high-energy limit. Our results may be utilised to derive the photonic decay rate of the Higgs-boson through next-to-next-to-leading order.}

\begin{document}

\maketitle

\newpage


\section{Introduction}

Since its discovery in 2012 at the Large Hadron Collider (LHC) by the two collaborations ATLAS and CMS \cite{Aad:2012tfa, Chatrchyan:2012ufa}, the Higgs-boson became one of the most promising candidates to study the Standard Model (SM) and physics beyond the SM. Even though the SM passed the most precise tests until now, small deviations between theoretical computations and experimental data could reveal missing pieces of a more complete theory of particle physics. It is therefore necessary to investigate the production and decay modes of the Higgs-boson in great detail. According to theory predictions for the branching-ratios (BR) Ref.~\cite{deFlorian:2016spz}, the decay of a 125 GeV Higgs-boson into a pair of bottom-quarks is favoured, but less significant for experimental studies due to the large background at hadron colliders. Despite the fact that the BR for the decay $H\to\gamma\gamma$ is of $\order{10^{-3}}$, Higgs-boson decay into a pair of photons belongs to the most relevant decay channels due to the high precision to which the final state particles can be measured.

Moreover, the feature that $H\to\gamma\gamma$ is a loop induced process makes it an appealing channel to determine not only the Higgs-boson mass with excellent resolution, but also to extract Yukawa couplings, since the Higgs-boson couples to all massive particles running in the loops.

Although the process at hand is loop induced and therefore hard to examine within the framework of a multi-loop calculation, the two-loop corrections to the Higgs-decay were computed a long time ago in the heavy-top limit in Refs.~\cite{Zheng:1990qa, Dawson:1992cy} and, subsequently, results covering the region below and even above the top-threshold followed with Refs.~\cite{Djouadi:1990aj,Melnikov:1993tj,Djouadi:1993ji,Spira:1995rr} via numerical integration. A decade later, these results became available in analytical form \cite{Fleischer:2004vb, Harlander:2005rq, Aglietti:2006tp}. However, the three-loop calculation seems to be more involved. Nevertheless, expansions in the regime, where the mass of the mediating quark is considered much larger than the mass of the Higgs-boson, have been employed to determine the three-loop form factor as a series expansion in terms of the fraction of the mentioned masses \cite{Steinhauser:1996wy, Maierhofer:2012vv}. The only analytical result currently available captures contributions originating from diagrams with one massless fermion loop \cite{Harlander:2019ioe}. Finally, the large logarithms of $\order{\alpha\alpha_s^2L^k}$ have been predicted in Refs.~\cite{Akhoury:2001mz, Anastasiou:2020vkr, Liu:2020tzd, Liu:2020wbn}.

The paper at hand was motivated by the authors of Refs.~\cite{Liu:2020tzd, Liu:2020wbn}, who kindly requested the availability of the Higgs-photon form factor expanded in the high-energy limit to perform consistency-checks with their own results. Since the diagrams that account for the Higgs-photon form factor form a subset of diagrams contributing to the Higgs-gluon form factor, we closely follow our previous publication Ref.~\cite{Czakon:2020vql}. Hence, the reduction table for the simplification to master integrals and their numerical solution, which was obtained via solving a system of differential equations, can be exploited to determine the desired expansions and the form factor itself.

\begin{figure}[t]
	\centering
    \begin{subfigure}[b]{.48\textwidth}
        \centering
        \includegraphics[scale=1.0]{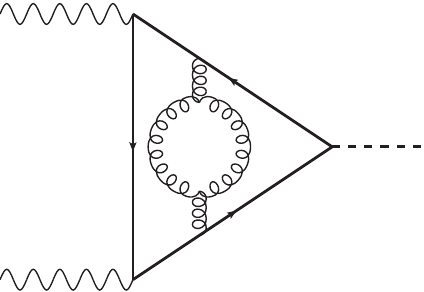}
        \caption{}
        \label{fig:feyn:a}
    \end{subfigure}%
        \begin{subfigure}[b]{.48\textwidth}
        \centering
        \includegraphics[scale=1.0]{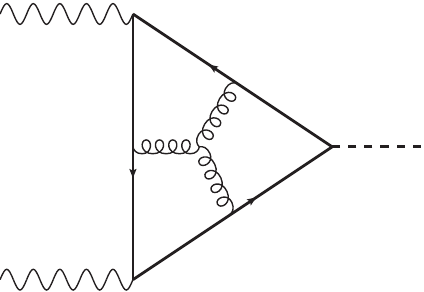}
        \caption{}
        \label{fig:feyn:b}
    \end{subfigure}%
    \\[18pt]
    \centering
    \begin{subfigure}[b]{.48\textwidth}
        \centering
        \includegraphics[scale=1.0]{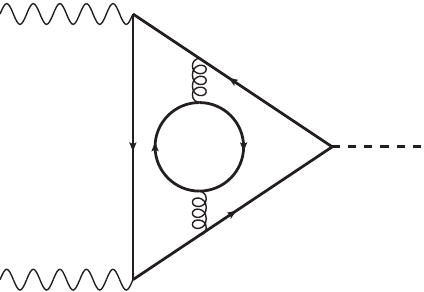}
        \caption{}
        \label{fig:feyn:c}
    \end{subfigure}%
        \begin{subfigure}[b]{.48\textwidth}
        \centering
        \includegraphics[scale=1.0]{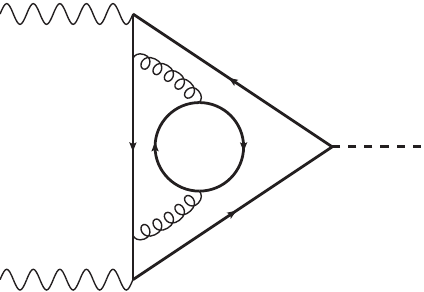}
        \caption{}
        \label{fig:feyn:d}
    \end{subfigure}%
    \\[18pt]
    \centering
    \begin{subfigure}[b]{.48\textwidth}
        \centering
        \includegraphics[scale=1.0]{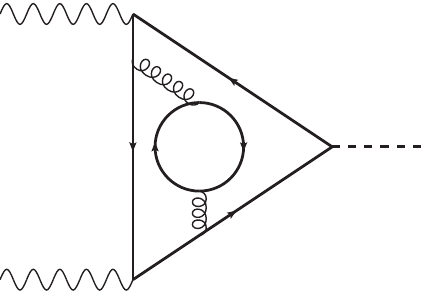}
        \caption{}
        \label{fig:feyn:e}
    \end{subfigure}%
    	\begin{subfigure}[b]{.48\textwidth}
        \centering
        \includegraphics[scale=1.0]{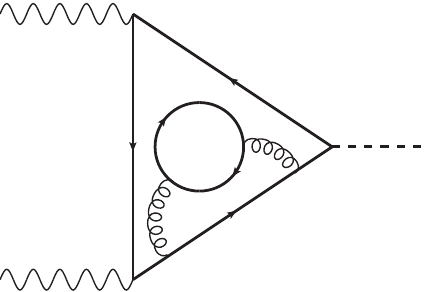}
        \caption{}
        \label{fig:feyn:f}
    \end{subfigure}%
    \\[18pt]
    \centering
    \begin{subfigure}[b]{.48\textwidth}
        \centering
        \includegraphics[scale=1.1]{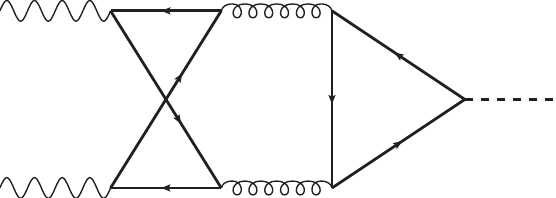}
        \caption{}
        \label{fig:feyn:g}
    \end{subfigure}%
    	\begin{subfigure}[b]{.48\textwidth}
        \centering
        \includegraphics[scale=1.1]{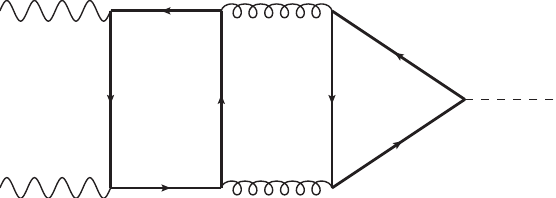}
        \caption{}
        \label{fig:feyn:h}
    \end{subfigure}%
\caption{Sample three-loop Feynman diagrams for the decay of a Higgs-boson into two photons. Diagrams \subref{fig:feyn:c}--\subref{fig:feyn:h} visualise all non-vanishing contributions involving two closed fermion loops.}
\label{fig:diags}
\end{figure}

Throughout this publication, we treat the diagrams shown in Fig.~\ref{fig:diags} that incorporate two fermion loops as follows: Either both fermions are massive quarks or one of them, in particular the one that couples to the Higgs-boson, is massive and the other one massless. In this way, we arrive at the three-loop Higgs-photon form factor in QCD with a single massive quark flavour.

This publication is structured as follows: In the following section, we clarify the notation and conventions used in this paper. Subsequently, we briefly discuss our findings and draw conclusions. Explicit results for the expansions of the missing piece of the three-loop form factor and information on the contents of the ancillary file are given in the appendices. An entire chapter dedicated for a thorough discussion on the technical details is given in Ref.~\cite{Czakon:2020vql}.
%


\section{Definitions}

In this section, we introduce the notation and conventions used throughout this paper. The process of interest is the decay of a Higgs-boson with arbitrary mass into two photons with momenta $p_1$ and $p_2$ and helicities $\lambda_1$ and $\lambda_2$. We write the amplitude as follows:
\begin{multline} \label{eq:amplitude}
-i \mathcal{M}\big[ H \to \gamma(p_1,\lambda_1) + \gamma(p_2,\lambda_2)
\big] \equiv
i \big[ (\epsilon_1 \cdot p_2) \, (\epsilon_2
\cdot p_1) - (\epsilon_1 \cdot \epsilon_2)  \, (p_2\cdot p_1) \big] \,
{1\over v} \frac{\alpha}{\pi} \, Q_q^2 \, \mathcal{C} \; .
\end{multline}
$Q_q$ denotes the electric charge of the top-quark, $\alpha$ is the electromagnetic coupling constant and $v$ indicates the Vacuum Expectation Value originating from the tree-level Lagrangian term $-M \bar{Q} Q H/v$, which is responsible for the coupling of the quark field to the Higgs-boson. For the photon polarisation vectors, the normalisation conditions hold:
\be
\epsilon_i \equiv \epsilon(\bm{p}_i,\lambda_i) \; , \qquad
\epsilon_i \cdot p_i = 0 \; , \qquad
\epsilon_i \cdot \epsilon^*_i = -1 \; , \qquad
i = 1,2 \; .
\ee
In accordance with Eq.~\eqref{eq:amplitude}, the {\bf Form Factor} $\mathcal{C}$ admits a perturbative expansion in terms of the strong coupling constant, $\alpha_s$.
\be \label{eq:expansion}
\begin{split}
\mathcal{C} &= \mathcal{C}^{(0)} + \frac{\alpha_s}\pi
\mathcal{C}^{(1)} + \left(\frac{\alpha_s}\pi\right)^2
\mathcal{C}^{(2)} + \order{\alpha_s^3} \; .
\end{split}
\ee
As far as renormalisation is concerned, we stick to the same conventions as in \cite{Czakon:2020vql} for the sake of convenience. We define the strong coupling constant in $\overline{\mathrm{MS}}$ scheme with massive-quark decoupling. The $\beta$-function for $n_l$ massless quarks gives rise to the dependence on the renormalisation scale: $\alpha_s \equiv \alpha_s^{(n_l)}(\mu)$. Furthermore, the quark-mass and henceforth the Yukawa coupling are renormalised in the on-shell scheme. All relevant constants for renormalisation and decoupling can be taken from \cite{Melnikov:2000qh, Melnikov:2000zc, Mitov:2006xs, Gray:1990yh, Bernreuther:1981sg}.

In contrast to the known one- and two-loop contributions, $\mathcal{C}^{(0)}$ and $\mathcal{C}^{(1)}$, respectively, the three-loop coefficient, $\mathcal{C}^{(2)}$, may be subdivided into contributions stemming from different classes of Feynman diagrams:
\be \label{eq:expansionC2}
\begin{split}
\mathcal{C}^{(2)} &= \mathcal{C}^{(2,0)} + n_h \, \mathcal{C}^{(2,1)}
+ n_l \, \mathcal{C}^{(2,2)}
+ \sum_{k=1}^{n_l} \left(\frac{Q_k}{Q_q}\right)^2 \, \mathcal{C}^{(2,3)} \; .
\end{split}
\ee
Here, the splitting into the four tree-loop coefficients, $\mathcal{C}^{(2,k)}$, is motivated by the fact that Feynman diagrams with more than one fermion loop contribute at three-loop level for the first time. $\mathcal{C}^{(2,0)}$ gathers all diagrams with exactly one closed fermion chain to which the external particles are necessarily attached. Two typical diagrams are shown in Figs.~\ref{fig:feyn:a}--\subref{fig:feyn:b}. Diagrams that contribute to $\mathcal{C}^{(2,1)}$ are those, which embed two massive fermion loops depicted in Figs.~\ref{fig:feyn:c}--\subref{fig:feyn:h}. We do not distinguish between diagrams in which one of the fermion loops is neither connected to the Higgs-boson nor to the external photons, as well as those where one of the fermion loops couples to the photons and the other one to the Higgs-boson. In this context, $n_h$ indicates the number of massive quarks not coupling to the Higgs-boson. With the three-loop coefficients $\mathcal{C}^{(2,2)}$ and $\mathcal{C}^{(2,3)}$, which are known analytically \cite{Harlander:2019ioe}, we associate all Feynman diagrams that involve one massless and one massive fermion loop. One usually differentiates between singlet and non-singlet contributions. Singlet diagrams (Figs.~\ref{fig:feyn:g}--\subref{fig:feyn:h}) collected in $\mathcal{C}^{(2,3)}$ incorporate one massive fermion loop attached to the Higgs-boson and one massless fermion loop that couples to the external photons. Hence, we have to sum over the electric charges of all massless fermion flavours. In contrast to that, the diagrams displayed in Figs.~\ref{fig:feyn:c}--\subref{fig:feyn:f} with a massless fermion loop in the centre account for the non-singlet part, $\mathcal{C}^{(2,2)}$. $\mathcal{C}^{(2,0)}$ encompasses non-singlet diagrams only, but as pointed out before, $\mathcal{C}^{(2,1)}$ covers both singlet and non-singlet parts.

The form factors and their individual components depend on the fraction of the masses of the Higgs-boson and the mediating massive quark and on the logarithm containing the renormalisation scale:
\begin{align}
\mathcal{C}^{(\ldots)} &\equiv \mathcal{C}^{(\ldots)}\left( z , \, L_\mu
\right) \; , \\[.2cm]
z \equiv \frac{s}{4M^2}+i0^+ \; , \qquad
L_\mu &\equiv \ln\left( -\frac{\mu^2}{s + i0^+} \right) \; , \qquad
s \equiv (p_1 + p_2)^2 \; .
\end{align}
In order to clarify the notation, we state the leading contribution:
\be
\mathcal{C}^{(0)} = {C_A\over T_F} {1\over 2z} \left\{ 1 -
  \left( 1 - {1\over z} \right) \left[ {1\over2} \ln\left(
      \frac{\sqrt{1-1/z}-1}{\sqrt{1-1/z}+1} \right) \right]^2 \right\}
\; ,
\ee
which in the heavy-top limit takes the value:
\be
\mathcal{C}^{(0)}\big[z = 0\big] = 2 \; .
\ee
The form factor is scale-independent implying
\be
\dv{\ln \mathcal{C}}{\ln \mu} = 0 \; .
\ee
Thus, the dependence of the form factor on the aforementioned logarithm, $L_\mu$, can be expressed with the aid of the coefficients of the QCD $\beta$-function:
\be \label{eq:scaleDependence}
\begin{aligned}
\mathcal{C}^{(0)} &= \mathcal{C}^{(0)}\big[ L_\mu = 0 \big] \; , \\[.2cm]
\mathcal{C}^{(1)} &= \mathcal{C}^{(1)}\big[ L_\mu = 0 \big] \; , \\[.2cm]
\mathcal{C}^{(2)} &= \mathcal{C}^{(2)}\big[ L_\mu = 0 \big] +
\frac{b_0}{4} \, \mathcal{C}^{(1)} \, L_\mu \; .
\end{aligned}
\ee
For the sake of completeness, we quote the first coefficient of the $\beta$-function:
\be
b_0 = \frac{11}{3} C_A - \frac{4}{3} T_F n_l \; .
\ee
%


\section{Results}

\begin{figure}[t]
\center
\includegraphics[width=\textwidth]{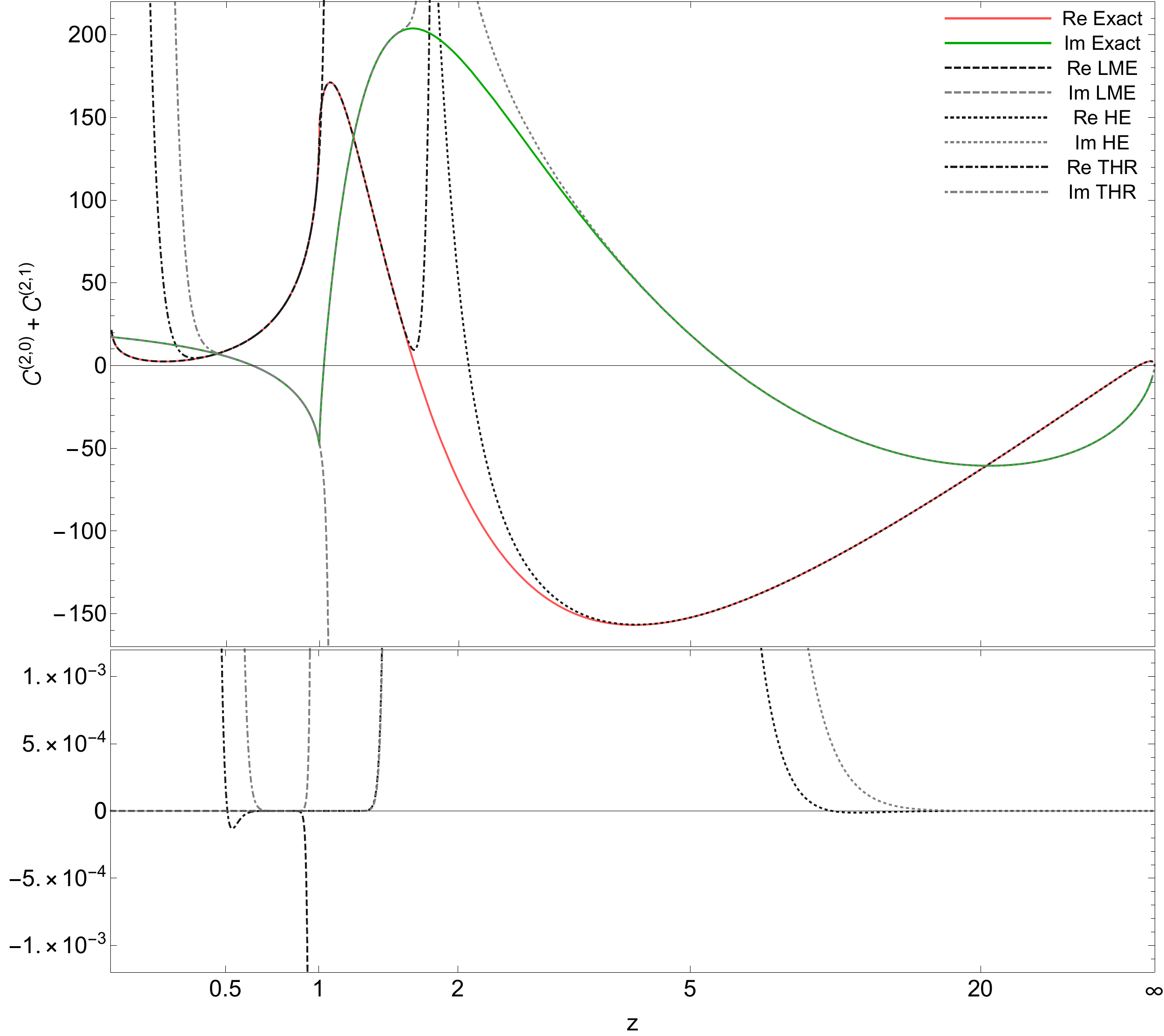}
\caption{Comparison of the large-mass expansion (LME) up to
  $\order{z^{100}}$, threshold expansion (THR) up to
  $\order{(1-z)^{20}}$ and high-energy expansion (HE) up to
  $\order{1/z^{8}}$ with the sum
	$\mathcal{C}^{(2,0)}+\mathcal{C}^{(2,1)}$ evaluated
	numerically ($L_\mu = 0$). The absolute difference
	between the exact result and the expansions is shown in the bottom panel.}
\label{fig:best}
\end{figure}
\begin{table}[t]
\center
\begin{tabular}{|c|c|c|c|}
\hline
$\rho$ & $\mathcal{C}^{(2,0)}+\mathcal{C}^{(2,1)}$ & $\rho$ & $\mathcal{C}^{(2,0)}+\mathcal{C}^{(2,1)}$ \\
\hline
$1/4$
 & $ 95.61412709+176.89801445 \, i $ & 
 $3/8$
 & $ -114.50707747+154.19306610 \, i $ \\ 
$51/200$
 & $ 83.07596079+184.20923088 \, i $ & 
 $19/50$
 & $ -118.45992102+149.95386054 \, i $ \\ 
$13/50$
 & $ 70.66622472+190.11797906 \, i $ & 
 $77/200$
 & $ -122.15870044+145.68482348 \, i $ \\ 
$53/200$
 & $ 58.47564245+194.78093076 \, i $ & 
 $39/100$
 & $ -125.61387922+141.39604695 \, i $ \\ 
$27/100$
 & $ 46.57137881+198.33693618 \, i $ & 
 $79/200$
 & $ -128.83557098+137.09659724 \, i $ \\ 
$11/40$
 & $ 35.00244955+200.90932597 \, i $ & 
 $2/5$
 & $ -131.83353409+132.79461434 \, i $ \\ 
$7/25$
 & $ 23.80378890+202.60783914 \, i $ & 
 $81/200$
 & $ -134.61717010+128.49740137 \, i $ \\ 
$57/200$
 & $ 12.99935281+203.53026304 \, i $ & 
 $41/100$
 & $ -137.19552522+124.21150499 \, i $ \\ 
$29/100$
 & $ 2.60451436+203.76384339 \, i $ & 
 $83/200$
 & $ -139.57729434+119.94278778 \, i $ \\ 
$59/200$
 & $ -7.37206987+203.38650552 \, i $ & 
 $21/50$
 & $ -141.77082686+115.69649353 \, i $ \\ 
$3/10$
 & $ -16.92699443+202.46791772 \, i $ & 
 $17/40$
 & $ -143.78413407+111.47730602 \, i $ \\ 
$61/200$
 & $ -26.06094755+201.07042088 \, i $ & 
 $43/100$
 & $ -145.62489760+107.28940212 \, i $ \\ 
$31/100$
 & $ -34.77779234+199.24984361 \, i $ & 
 $87/200$
 & $ -147.30047872+103.13649972 \, i $ \\ 
$63/200$
 & $ -43.08383477+197.05621891 \, i $ & 
 $11/25$
 & $ -148.81792823+99.02190109 \, i $ \\ 
$8/25$
 & $ -50.98723878+194.53441538 \, i $ & 
 $89/200$
 & $ -150.18399666+94.94853204 \, i $ \\ 
$13/40$
 & $ -58.49755811+191.72469437 \, i $ & 
 $9/20$
 & $ -151.40514481+90.91897739 \, i $ \\ 
$33/100$
 & $ -65.62536110+188.66320244 \, i $ & 
 $91/200$
 & $ -152.48755434+86.93551306 \, i $ \\ 
$67/200$
 & $ -72.38193010+185.38240734 \, i $ & 
 $23/50$
 & $ -153.43713829+83.00013522 \, i $ \\ 
$17/50$
 & $ -78.77902071+181.91148449 \, i $ & 
 $93/200$
 & $ -154.25955161+79.11458662 \, i $ \\ 
$69/200$
 & $ -84.82866943+178.27666003 \, i $ & 
 $47/100$
 & $ -154.96020144+75.28038056 \, i $ \\ 
$7/20$
 & $ -90.54304032+174.50151571 \, i $ & 
 $19/40$
 & $ -155.54425729+71.49882262 \, i $ \\ 
$71/200$
 & $ -95.93430346+170.60726020 \, i $ & 
 $12/25$
 & $ -156.01666080+67.77103038 \, i $ \\ 
$9/25$
 & $ -101.01453909+166.61297072 \, i $ & 
 $97/200$
 & $ -156.38213534+64.09795135 \, i $ \\ 
$73/200$
 & $ -105.79566262+162.53580859 \, i $ & 
 $49/100$
 & $ -156.64519521+60.48037931 \, i $ \\ 
$37/100$
 & $ -110.28936677+158.39121163 \, i $ & 
 $99/200$
 & $ -156.81015439+56.91896909 \, i $ \\ 

\hline
\end{tabular}
\caption{Numerical values of the three-loop coefficient $\mathcal{C}^{(2,0)}+\mathcal{C}^{(2,1)}$ at
$L_\mu = 0$, for $1/4 \leq \rho \equiv z/(4+z) < 1/2$.}
\label{tab:num1}
\end{table}
\begin{table}[t]
\center
\begin{tabular}{|c|c|c|c|}
\hline
$\rho$ & $\mathcal{C}^{(2,0)}+\mathcal{C}^{(2,1)}$ & $\rho$ & $\mathcal{C}^{(2,0)}+\mathcal{C}^{(2,1)}$ \\
\hline
$1/2$
 & $ -156.88113479+53.41425013 \, i $ & 
 $5/8$
 & $ -137.132840055-15.702661678 \, i $ \\ 
$101/200$
 & $ -156.86207296+49.96663944 \, i $ & 
 $63/100$
 & $ -135.749456670-17.756266155 \, i $ \\ 
$51/100$
 & $ -156.75672881+46.57645720 \, i $ & 
 $127/200$
 & $ -134.335469634-19.758491416 \, i $ \\ 
$103/200$
 & $ -156.56869815+43.24393631 \, i $ & 
 $16/25$
 & $ -132.892081100-21.709726182 \, i $ \\ 
$13/25$
 & $ -156.30142058+39.96922935 \, i $ & 
 $129/200$
 & $ -131.420442234-23.610352498 \, i $ \\ 
$21/40$
 & $ -155.95818634+36.75241608 \, i $ & 
 $13/20$
 & $ -129.921655363-25.460744615 \, i $ \\ 
$53/100$
 & $ -155.54214291+33.59351066 \, i $ & 
 $131/200$
 & $ -128.396776022-27.261267930 \, i $ \\ 
$107/200$
 & $ -155.05630130+30.49246836 \, i $ & 
 $33/50$
 & $ -126.846814904-29.012277977 \, i $ \\ 
$27/50$
 & $ -154.50354211+27.44919179 \, i $ & 
 $133/200$
 & $ -125.272739719-30.714119456 \, i $ \\ 
$109/200$
 & $ -153.88662135+24.46353662 \, i $ & 
 $67/100$
 & $ -123.675476970-32.367125304 \, i $ \\ 
$11/20$
 & $ -153.20817593+21.53531684 \, i $ & 
 $27/40$
 & $ -122.055913644-33.971615786 \, i $ \\ 
$111/200$
 & $ -152.47072896+18.66430965 \, i $ & 
 $17/25$
 & $ -120.414898822-35.527897609 \, i $ \\ 
$14/25$
 & $ -151.67669487+15.85025982 \, i $ & 
 $137/200$
 & $ -118.753245223-37.036263057 \, i $ \\ 
$113/200$
 & $ -150.82838414+13.09288386 \, i $ & 
 $69/100$
 & $ -117.071730670-38.496989120 \, i $ \\ 
$57/100$
 & $ -149.92800802+10.39187374 \, i $ & 
 $139/200$
 & $ -115.371099493-39.910336636 \, i $ \\ 
$23/40$
 & $ -148.97768284+7.74690031 \, i $ & 
 $7/10$
 & $ -113.652063866-41.276549423 \, i $ \\ 
$29/50$
 & $ -147.97943427+5.15761652 \, i $ & 
 $141/200$
 & $ -111.915305089-42.595853392 \, i $ \\ 
$117/200$
 & $ -146.93520129+2.62366027 \, i $ & 
 $71/100$
 & $ -110.161474804-43.868455656 \, i $ \\ 
$59/100$
 & $ -145.84684003+0.14465717 \, i $ & 
 $143/200$
 & $ -108.391196166-45.094543592 \, i $ \\ 
$119/200$
 & $ -144.71612740-2.27977708 \, i $ & 
 $18/25$
 & $ -106.605064963-46.274283888 \, i $ \\ 
$3/5$
 & $ -143.54476456-4.65003435 \, i $ & 
 $29/40$
 & $ -104.803650678-47.407821537 \, i $ \\ 
$121/200$
 & $ -142.33438025-6.96651202 \, i $ & 
 $73/100$
 & $ -102.987497519-48.495278782 \, i $ \\ 
$61/100$
 & $ -141.086533913-9.229611000 \, i $ & 
 $147/200$
 & $ -101.157125397-49.536754007 \, i $ \\ 
$123/200$
 & $ -139.802718709-11.439733991 \, i $ & 
 $37/50$
 & $ -99.313030875-50.532320551 \, i $ \\ 
$31/50$
 & $ -138.484364399-13.597283772 \, i $ & 
 $149/200$
 & $ -97.455688067-51.482025454 \, i $ \\ 
$5/8$
 & $ -137.132840055-15.702661678 \, i $ & 
 $3/4$
 & $ -95.585549514-52.385888101 \, i $ \\ 

\hline
\end{tabular}
\caption{Numerical values of the three-loop coefficient $\mathcal{C}^{(2,0)}+\mathcal{C}^{(2,1)}$ at
$L_\mu = 0$, for $1/2 \leq \rho \equiv z/(4+z) \leq 3/4$.}
\label{tab:num2}
\end{table}
In this section, we briefly present out findings. The scale-dependence is fixed such that $L_\mu = 0$ and can easily be restored by applying Eqs.~\eqref{eq:scaleDependence}.

We check our results for the light-fermion contributions, in particular, the three-loop coefficients $\mathcal{C}^{(2,2)}$ and $\mathcal{C}^{(2,3)}$, numerically against the analytical results in Ref.~\cite{Harlander:2019ioe}. For the numerical probes as well as for the expansions in the kinematic limits, we find full agreement.

Similar to our previous work, the exact result for the Higgs-photon form factor at three-loop level, $\mathcal{C}^{(2)}$, is stored in the form of a univariate interpolation based on nearly 200.000 numerical probes distributed over the physical parameter space in the variable $z$. Other than that, we derived high-order large-mass, threshold and high-energy expansions, which cover most parts of the parameter space to sufficient precision. The radii of convergence of the three expansions are limited due to singularities located at $z=0$, $z=1$ and $1/z=0$. The ancillary file \cite{mathfile} shipped with this publication contains the large-mass expansion with exact coefficients truncated at $\order{z^{100}}$, the threshold expansion truncated at $\order{(1-z)^{20}}$ and the high-energy expansion truncated at $\order{1/z^{8}}$. The latter ones are expansions with numerical coefficients. We choose the truncation order of the numerical expansions such that we can confidently guarantee the correctness of at least ten digits for every numerical coefficient.

A comparison of the mentioned expansions with the exact numerical result for the sum of three-loop coefficients $\mathcal{C}^{(2,0)}+\mathcal{C}^{(2,1)}$ is illustrated in Fig.~\ref{fig:best}. For those values of $z$ which are not covered by expansions, we provide interpolation tables in Tabs.~\ref{tab:num1}~and~\ref{tab:num2}, where we reshaped the domain of positive $z$ values to the interval $(0,1)$ by applying the conformal mapping
\be
z(\rho) \equiv \frac{4\rho}{1 - \rho} \; , \qquad
\rho(z) = \frac{z}{4+z} \; , \qquad
\rho \in (0,1) \; .
\ee
With a relative error of at most $10^{-5}$, the exact result for the sum of three-loop coefficients $\mathcal{C}^{(2,0)}+\mathcal{C}^{(2,1)}$ is approximated as follows:
\begin{center}
\begin{tabular}{ll}
$0 < \rho < 1/6$ & - large-mass expansion, Appendix~\ref{sec:LME} and Fig.~\ref{fig:LME}; \\[.2cm]
$1/6 \leq \rho < 1/4$ & - threshold expansion, Appendix~\ref{sec:THR} and Fig.~\ref{fig:THR}; \\[.2cm]
$1/4 \leq \rho < 3/4$ & - interpolation of a sample of numerical values, Tabs.~\ref{tab:num1}~and~\ref{tab:num2}; \\[.2cm]
$3/4 \leq \rho < 1$ & - high-energy expansion, Appendix~\ref{sec:HE} and Fig.~\ref{fig:HE}.
\end{tabular}
\end{center}
%


\section{Conclusions and outlook}

Provided the findings of this paper, the Higgs-photon form factor is now known exactly at the three-loop level in QCD with a single massive quark-flavour. Moreover, the longing for the desired expansions has been satisfied. We presented the results with the Yukawa coupling renormalised in the on-shell scheme, which can be translated to any other scheme due to the fact that the one- and two-loop results are available in analytical form.

We finally note that our results may be utilised to obtain the cross section for Higgs-boson production via photon-photon fusion and the photonic decay rate of a Higgs-boson through next-to-next-to-leading order in QCD.

Let us again emphasise that the form factor with the most general quark-mass dependence requires additional elaboration of the diagrams with two closed fermion chains. We postpone this analysis to future publications.


\section*{Acknowledgements}

This work was supported by the Deutsche Forschungsgemeinschaft (DFG) under grants 396021762 - TRR 257 and 400140256 - GRK 2497: The physics of the heaviest particles at the Large Hardon Collider.


\newpage

\appendix


\section{Large-mass expansion} \label{sec:LME}

\begin{figure}[t]
\center
\includegraphics[width=\textwidth]{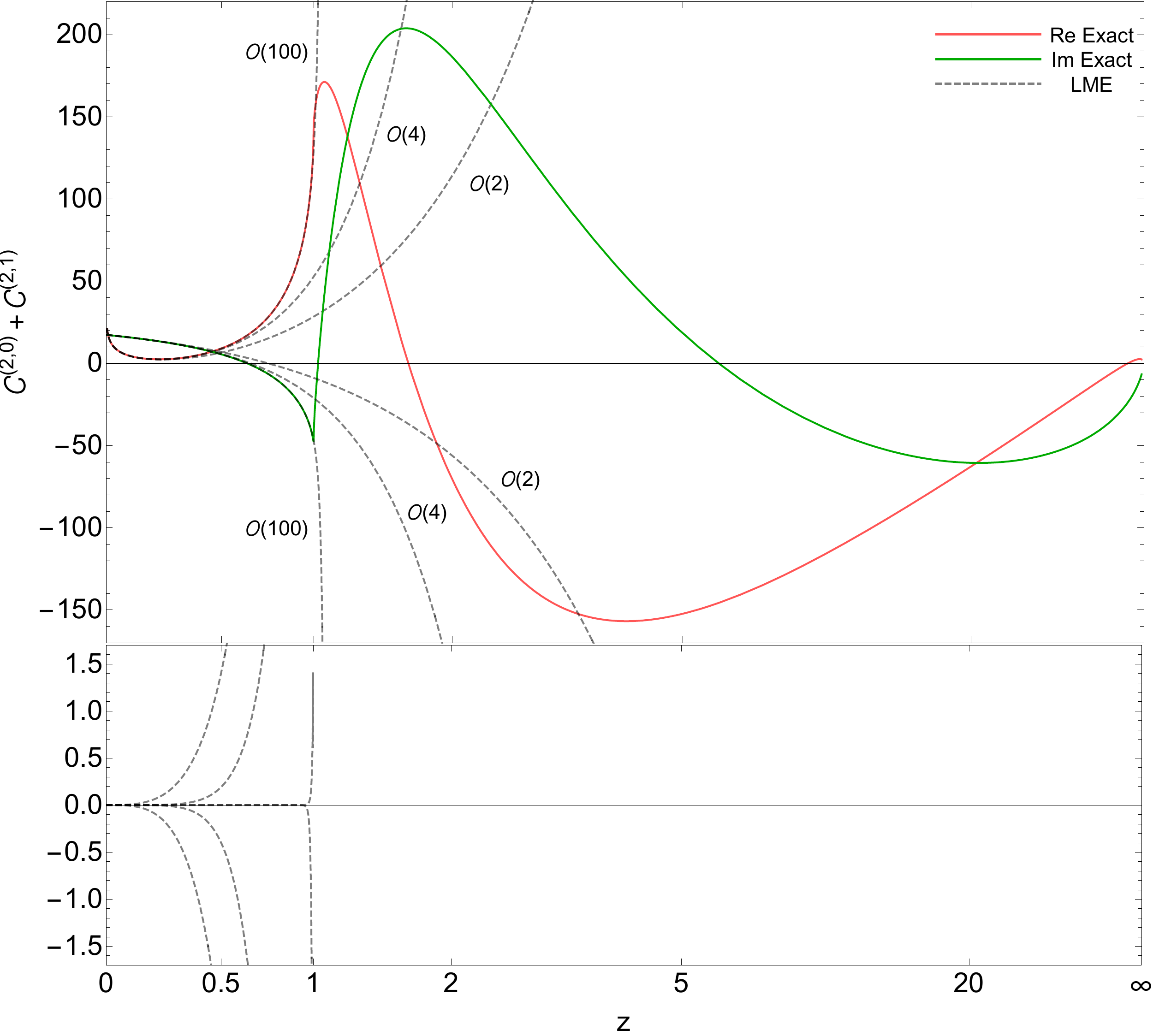}
\caption{Comparison of the large-mass expansion (LME) up to $\order{z^{2}}$, $\order{z^{4}}$ and $\order{z^{100}}$ with the sum $\mathcal{C}^{(2,0)}+\mathcal{C}^{(2,1)}$ evaluated numerically ($L_\mu = 0)$. The absolute difference between the exact result and the expansions is shown in the bottom panel.}
\label{fig:LME}
\end{figure}
\be
C^{(2,0)}+C^{(2,1)} = \sum_{n=0}^\infty \left( a_{n,0} + a_{n,1} \, L_s\right) \, z^n
\; , \qquad
L_s \equiv \ln\left( -\frac{s}{M^2} - i0^+ \right) \; ,
\ee
\begin{align}
\begin{autobreak}
\mathcal{C}^{(2,0)}+\mathcal{C}^{(2,1)} =
-1.777777778
-5.500000000 \, L_s
+\Big(13.11710090
+4.970370370 \, L_s\Big) \, z
+\Big(13.06368527
+3.350405644 \, L_s\Big) \, z^{2}
+\Big(10.46030710
+2.278524565 \, L_s\Big) \, z^{3}
+\Big(8.320460471
+1.610008230 \, L_s\Big) \, z^{4}
+\Big(6.703047617
+1.188440003 \, L_s\Big) \, z^{5}
+\Big(5.493827219
+0.9141819603 \, L_s\Big) \, z^{6}
+\Big(4.588959115
+0.7191908041 \, L_s\Big) \, z^{7}
+\Big(3.885458574
+0.5838547944 \, L_s\Big) \, z^{8}
+\Big(3.340520295
+0.4795265645 \, L_s\Big) \, z^{9}
+\Big(2.900355777
+0.4036136804 \, L_s\Big) \, z^{10}
+\Big(2.548334424
+0.3416903422 \, L_s\Big) \, z^{11}
+\Big(2.255192075
+0.2950940542 \, L_s\Big) \, z^{12}
+\Big(2.014632209
+0.2554585379 \, L_s\Big) \, z^{13}
+\Big(1.809436750
+0.2248978875 \, L_s\Big) \, z^{14}
+\Big(1.637551699
+0.1980436051 \, L_s\Big) \, z^{15}
+\Big(1.488094722
+0.1769594650 \, L_s\Big) \, z^{16}
+\Big(1.360807921
+0.1579418622 \, L_s\Big) \, z^{17}
+\Big(1.248390264
+0.1428040226 \, L_s\Big) \, z^{18}
+\Big(1.151340958
+0.1288517530 \, L_s\Big) \, z^{19}
+\Big(1.064516424
+0.1176276537 \, L_s\Big) \, z^{20}
+\Big(0.9887120593
+0.1070919600 \, L_s\Big) \, z^{21}
+\Big(0.9201563781
+0.09854608319 \, L_s\Big) \, z^{22}
+\Big(0.8597314175
+0.09039720912 \, L_s\Big) \, z^{23}
+\Big(0.8045792128
+0.08374429753 \, L_s\Big) \, z^{24}
+\Big(0.7555733042
+0.07731261247 \, L_s\Big) \, z^{25}
+\Big(0.7104879787
+0.07203463456 \, L_s\Big) \, z^{26}
+\Big(0.6701469601
+0.06686962473 \, L_s\Big) \, z^{27}
+\Big(0.6327767878
+0.06261386339 \, L_s\Big) \, z^{28}
+\Big(0.5991358278
+0.05840349687 \, L_s\Big) \, z^{29}
+\Big(0.5677833750
+0.05492318564 \, L_s\Big) \, z^{30}
+\Big(0.5394092405
+0.05144585116 \, L_s\Big) \, z^{31}
+\Big(0.5128235733
+0.04856423129 \, L_s\Big) \, z^{32}
+\Big(0.4886501782
+0.04565905067 \, L_s\Big) \, z^{33}
+\Big(0.4658923404
+0.04324686803 \, L_s\Big) \, z^{34}
+\Big(0.4451128638
+0.04079477885 \, L_s\Big) \, z^{35}
+\Big(0.4254664886
+0.03875577657 \, L_s\Big) \, z^{36}
+\Big(0.4074608421
+0.03666714389 \, L_s\Big) \, z^{37}
+\Big(0.3903713631
+0.03492846512 \, L_s\Big) \, z^{38}
+\Big(0.3746563443
+0.03313477019 \, L_s\Big) \, z^{39}
+\Big(0.3596887907
+0.03164046183 \, L_s\Big) \, z^{40}
+\Big(0.3458831478
+0.03008860435 \, L_s\Big) \, z^{41}
+\Big(0.3326922897
+0.02879512859 \, L_s\Big) \, z^{42}
+\Big(0.3204917987
+0.02744344429 \, L_s\Big) \, z^{43}
+\Big(0.3088006713
+0.02631651964 \, L_s\Big) \, z^{44}
+\Big(0.2979600688
+0.02513194782 \, L_s\Big) \, z^{45}
+\Big(0.2875442851
+0.02414428832 \, L_s\Big) \, z^{46}
+\Big(0.2778639781
+0.02310032496 \, L_s\Big) \, z^{47}
+\Big(0.2685401083
+0.02222999757 \, L_s\Big) \, z^{48}
+\Big(0.2598562662
+0.02130519471 \, L_s\Big) \, z^{49}
+\Big(0.2514731205
+0.02053440900 \, L_s\Big) \, z^{50}
+\Big(0.2436501982
+0.01971125995 \, L_s\Big) \, z^{51}
+\Big(0.2360822089
+0.01902546512 \, L_s\Big) \, z^{52}
+\Big(0.2290072400
+0.01828956618 \, L_s\Big) \, z^{53}
+\Big(0.2221493831
+0.01767676928 \, L_s\Big) \, z^{54}
+\Big(0.2157275788
+0.01701618233 \, L_s\Big) \, z^{55}
+\Big(0.2094914381
+0.01646642798 \, L_s\Big) \, z^{56}
+\Big(0.2036427601
+0.01587119071 \, L_s\Big) \, z^{57}
+\Big(0.1979534480
+0.01537616013 \, L_s\Big) \, z^{58}
+\Big(0.1926099161
+0.01483790614 \, L_s\Big) \, z^{59}
+\Big(0.1874036433
+0.01439060512 \, L_s\Big) \, z^{60}
+\Big(0.1825072020
+0.01390226658 \, L_s\Big) \, z^{61}
+\Big(0.1777293453
+0.01349678066 \, L_s\Big) \, z^{62}
+\Big(0.1732301604
+0.01305235368 \, L_s\Big) \, z^{63}
+\Big(0.1688337129
+0.01268365449 \, L_s\Big) \, z^{64}
+\Big(0.1646888013
+0.01227801247 \, L_s\Big) \, z^{65}
+\Big(0.1606331223
+0.01194180331 \, L_s\Big) \, z^{66}
+\Big(0.1568052431
+0.01157054741 \, L_s\Big) \, z^{67}
+\Big(0.1530550417
+0.01126313939 \, L_s\Big) \, z^{68}
+\Big(0.1495117915
+0.01092247795 \, L_s\Big) \, z^{69}
+\Big(0.1460362962
+0.01064068979 \, L_s\Big) \, z^{70}
+\Big(0.1427493678
+0.01032734039 \, L_s\Big) \, z^{71}
+\Big(0.1395216432
+0.01006841703 \, L_s\Big) \, z^{72}
+\Big(0.1364662142
+0.009779526526 \, L_s\Big) \, z^{73}
+\Big(0.1334625959
+0.009541072480 \, L_s\Big) \, z^{74}
+\Big(0.1306168216
+0.009274151328 \, L_s\Big) \, z^{75}
+\Big(0.1278164469
+0.009054075932 \, L_s\Big) \, z^{76}
+\Big(0.1251610381
+0.008806943918 \, L_s\Big) \, z^{77}
+\Big(0.1225454532
+0.008603416011 \, L_s\Big) \, z^{78}
+\Big(0.1200633223
+0.008374157224 \, L_s\Big) \, z^{79}
+\Big(0.1176161532
+0.008185567502 \, L_s\Big) \, z^{80}
+\Big(0.1152921168
+0.007972492768 \, L_s\Big) \, z^{81}
+\Big(0.1129987915
+0.007797422353 \, L_s\Big) \, z^{82}
+\Big(0.1108193188
+0.007599037760 \, L_s\Big) \, z^{83}
+\Big(0.1086668325
+0.007436231834 \, L_s\Big) \, z^{84}
+\Big(0.1066198321
+0.007251212232 \, L_s\Big) \, z^{85}
+\Big(0.1045965469
+0.007099557830 \, L_s\Big) \, z^{86}
+\Big(0.1026711854
+0.006926724446 \, L_s\Big) \, z^{87}
+\Big(0.1007666602
+0.006785231676 \, L_s\Big) \, z^{88}
+\Big(0.09895320635
+0.006623533114 \, L_s\Big) \, z^{89}
+\Big(0.09715805042
+0.006491319205 \, L_s\Big) \, z^{90}
+\Big(0.09544774254
+0.006339815263 \, L_s\Big) \, z^{91}
+\Big(0.09375349018
+0.006216090979 \, L_s\Big) \, z^{92}
+\Big(0.09213842100
+0.006073938819 \, L_s\Big) \, z^{93}
+\Big(0.09053742317
+0.005957996834 \, L_s\Big) \, z^{94}
+\Big(0.08901044095
+0.005824439116 \, L_s\Big) \, z^{95}
+\Big(0.08749577159
+0.005715644039 \, L_s\Big) \, z^{96}
+\Big(0.08605039426
+0.005589998712 \, L_s\Big) \, z^{97}
+\Big(0.08461576903
+0.005487778501 \, L_s\Big) \, z^{98}
+\Big(0.08324610956
+0.005369429986 \, L_s\Big) \, z^{99}
+\Big(0.08188581505
+0.005273268529 \, L_s\Big) \, z^{100}

+ \order{z^{101}} \; .
\end{autobreak}
\end{align}
In order to present the high-order LME in a space-saving way, we refrain from showing the analytical result here and refer to Ref.~\cite{mathfile} for the exact expansion. After conversion to a quark-mass renormalised in $\overline{\mathrm{MS}}$ scheme, we find full agreement with Ref.~\cite{Maierhofer:2012vv} up to $\order{z^{20}}$.


\newpage

\section{Threshold expansion} \label{sec:THR}

\begin{figure}[t]
\center
\includegraphics[width=\textwidth]{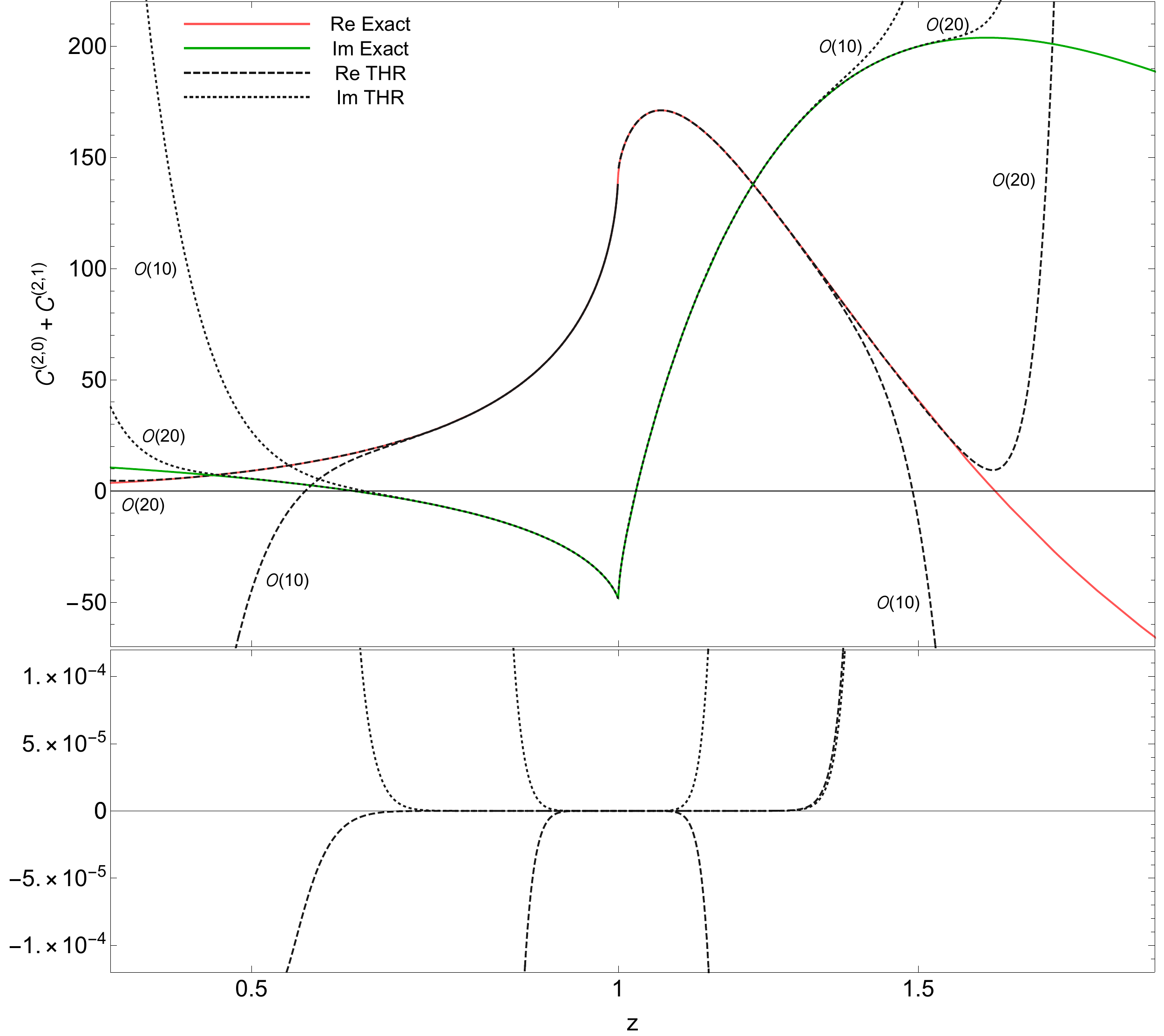}
\caption{Comparison of the threshold expansion (THR) up to $\order{(1-z)^{10}}$ and $\order{(1-z)^{20}}$ with the sum $\mathcal{C}^{(2,0)}+\mathcal{C}^{(2,1)}$ evaluated numerically ($L_\mu = 0$). The absolute difference between the exact result and the expansions is shown in the bottom panel.}
\label{fig:THR}
\end{figure}
\be
\begin{aligned}
&C^{(2,0)}+C^{(2,1)} = \sum_{n=0}^\infty \left( b_{n,0} + b_{n,1} \, L_t +
  b_{n,2} \, L_t^2 \right) \, t^n \; , \\[.2cm]
&L_t \equiv \ln\left( 1 - z \right) \; , \qquad t \equiv
\sqrt{1 - z} = \exp(L_t/2) \; ,
\end{aligned}
\ee
\begin{align}
\begin{autobreak}
\mathcal{C}^{(2,0)}+\mathcal{C}^{(2,1)} =
143.7364241
-48.71649342 \, i 
-177.3504510 \, t 
+\Big(197.6561763
-204.1570374 \, i
+(148.1633930
-170.5345217 \, i) \, L_t 
-27.14141210 \, L_t^2\Big) \, t^{2}
+\Big(53.77375045
 +59.37899426 \, i
-172.2570927 \, L_t\Big) \, t^{3}
+\Big(-2182.218552
 +696.9359899 \, i
+(70.58782969
-113.6896812 \, i) \, L_t 
-72.92541030 \, L_t^2\Big) \, t^{4}
+\Big(2592.396897
-775.0446867 \, i
+(-940.1458201
 +289.5083958 \, i) \, L_t 
+92.15338451 \, L_t^2\Big) \, t^{5}
+\Big(-6442.015967
 +2189.501187 \, i
+(-156.2174980
 +79.58277681 \, i) \, L_t 
-56.78677940 \, L_t^2\Big) \, t^{6}
+\Big(7101.227732
-2217.403057 \, i
+(-2281.118997
 +800.9732283 \, i) \, L_t 
+314.5316023 \, L_t^2\Big) \, t^{7}
+\Big(-12280.42300
 +4176.216119 \, i
+(-416.8534109
 +376.8000861 \, i) \, L_t 
+38.55934333 \, L_t^2\Big) \, t^{8}
+\Big(13282.39833
-4186.062962 \, i
+(-4084.925834
 +1495.448725 \, i) \, L_t 
+676.5815942 \, L_t^2\Big) \, t^{9}
+\Big(-19513.93758
 +6637.073703 \, i
+(-634.2474597
 +760.6381049 \, i) \, L_t 
+219.9642388 \, L_t^2\Big) \, t^{10}
+\Big(20956.86608
-6661.423781 \, i
+(-6245.914043
 +2348.223277 \, i) \, L_t 
+1181.316740 \, L_t^2\Big) \, t^{11}
+\Big(-28019.96998
 +9579.038585 \, i
+(-751.3909196
 +1220.072379 \, i) \, L_t 
+490.9776099 \, L_t^2\Big) \, t^{12}
+\Big(30014.75264
-9648.580578 \, i
+(-8657.469901
 +3342.107976 \, i) \, L_t 
+1829.712389 \, L_t^2\Big) \, t^{13}
+\Big(-37710.95435
 +13020.79214 \, i
+(-722.4485167
 +1747.349447 \, i) \, L_t 
+853.7398177 \, L_t^2\Big) \, t^{14}
+\Big(40390.07403
-13164.14293 \, i
+(-11201.33327
 +4464.349253 \, i) \, L_t 
+2621.968191 \, L_t^2\Big) \, t^{15}
+\Big(-48521.24097
 +16986.69106 \, i
+(-508.4575273
 +2336.654212 \, i) \, L_t 
+1309.673614 \, L_t^2\Big) \, t^{16}
+\Big(52049.17759
-17230.58951 \, i
+(-13736.34303
 +5705.036736 \, i) \, L_t 
+3557.958838 \, L_t^2\Big) \, t^{17}
+\Big(-60399.79929
 +21503.96892 \, i
+(-75.04397778
 +2983.425813 \, i) \, L_t 
+1859.791012 \, L_t^2\Big) \, t^{18}
+\Big(64986.27918
-21873.52288 \, i
+(-16085.54882
 +7056.200309 \, i) \, L_t 
+4637.418784 \, L_t^2\Big) \, t^{19}
+\Big(-73305.84110
 +26601.30209 \, i
+(608.8980624
 +3683.967554 \, i) \, L_t 
+2504.848471 \, L_t^2\Big) \, t^{20}
+\Big(79223.34316
-27120.21267 \, i
+(-18020.18143
 +8511.258872 \, i) \, L_t 
+5860.025054 \, L_t^2\Big) \, t^{21}
+\Big(-87206.03702
 +32308.02280 \, i
+(1572.025947
 +4435.206855 \, i) \, L_t 
+3245.433200 \, L_t^2\Big) \, t^{22}
+\Big(94813.29712
-32998.77290 \, i
+(-19238.78222
 +10064.66381 \, i) \, L_t 
+7225.436525 \, L_t^2\Big) \, t^{23}
+\Big(-102072.6614
 +38653.66682 \, i
+(2841.087074
 +5234.538704 \, i) \, L_t 
+4082.014739 \, L_t^2\Big) \, t^{24}
+\Big(111846.4424
-39537.67660 \, i
+(-19339.41674
 +11711.65750 \, i) \, L_t 
+8733.313071 \, L_t^2\Big) \, t^{25}
+\Big(-117882.3077
 +45667.70560 \, i
+(4441.290158
 +6079.718781 \, i) \, L_t 
+5014.977594 \, L_t^2\Big) \, t^{26}
+\Big(130460.4625
-46765.46043 \, i
+(-17782.29142
 +13448.10353 \, i) \, L_t 
+10383.32486 \, L_t^2\Big) \, t^{27}
+\Big(-134614.9707
 +53379.38527 \, i
+(6396.571505
 +6968.787760 \, i) \, L_t 
+6044.642843 \, L_t^2\Big) \, t^{28}
+\Big(150854.8781
-54710.54167 \, i
+(-13839.21226
 +15270.36352 \, i) \, L_t 
+12175.15662 \, L_t^2\Big) \, t^{29}
+\Big(-152253.3725
 +61817.62989 \, i
+(8729.792470
 +7900.016030 \, i) \, L_t 
+7171.282997 \, L_t^2\Big) \, t^{30}
+\Big(173311.2659
-63401.10391 \, i
+(-6525.100034
 +17175.20536 \, i) \, L_t 
+14108.50932 \, L_t^2\Big) \, t^{31}
+\Big(-170782.4563
 +71010.98452 \, i
+(11462.89010
 +8871.862283 \, i) \, L_t 
+8395.132546 \, L_t^2\Big) \, t^{32}
+\Big(198221.1192
-72865.02611 \, i
+(5494.900238
 +19159.73344 \, i) \, L_t 
+16183.10042 \, L_t^2\Big) \, t^{33}
+\Big(-190188.9992
 +80987.58341 \, i
+(14616.99489
 +9882.941815 \, i) \, L_t 
+9716.395645 \, L_t^2\Big) \, t^{34}
+\Big(226123.9546
-83129.83963 \, i
+(24031.45904
 +21221.33438 \, i) \, L_t 
+18398.66346 \, L_t^2\Big) \, t^{35}
+\Big(-210461.3098
 +91775.13446 \, i
+(18212.52478
 +10932.00180 \, i) \, L_t 
+11135.25185 \, L_t^2\Big) \, t^{36}
+\Big(257759.2233
-94222.70360 \, i
+(51542.09097
 +23357.63429 \, i) \, L_t 
+20754.94726 \, L_t^2\Big) \, t^{37}
+\Big(-231588.9897
 +103400.9142 \, i
+(22269.26158
 +12017.90168 \, i) \, L_t 
+12651.86050 \, L_t^2\Big) \, t^{38}
+\Big(294136.8713
-106170.3924 \, i
+(91363.20676
 +25566.46448 \, i) \, L_t 
+23251.71496 \, L_t^2\Big) \, t^{39}
+\Big(-253562.7417
 +115891.7695 \, i
+(26806.41403
 +13139.59741 \, i) \, L_t 
+14266.36404 \, L_t^2\Big) \, t^{40}

+ \order{t^{41}} \; .
\end{autobreak}
\end{align}
%


\newpage

\section{High-energy expansion} \label{sec:HE}

\begin{figure}[t]
\center
\includegraphics[width=\textwidth]{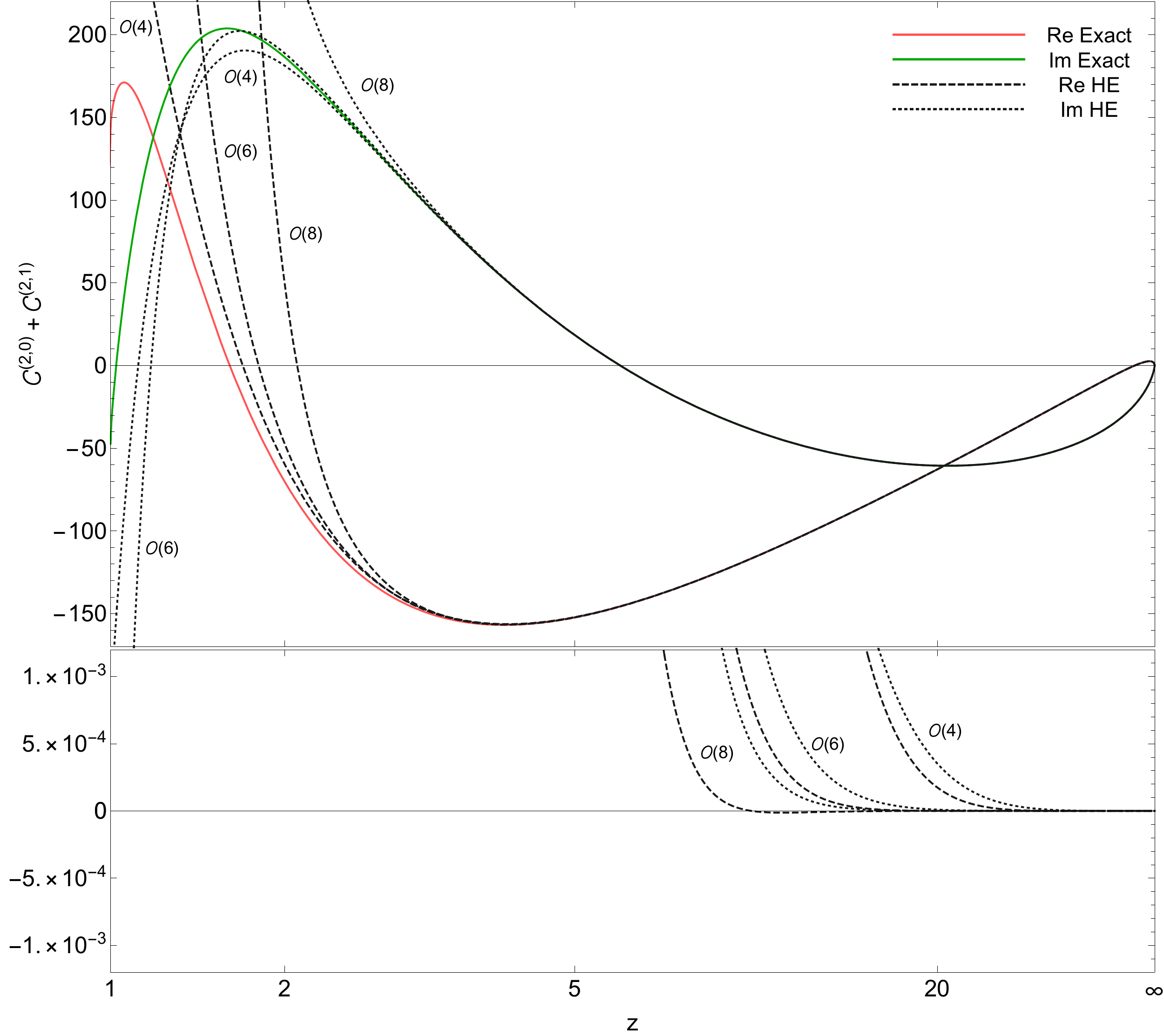}
\caption{Comparison of the high-energy expansion (HE) up to $\order{1/z^{4}}$, $\order{1/z^{6}}$ and $\order{1/z^{8}}$ with the sum $\mathcal{C}^{(2,0)}+\mathcal{C}^{(2,1)}$ evaluated numerically ($L_\mu = 0$). The absolute difference between the exact result and the expansions is shown in the bottom panel.}
\label{fig:HE}
\end{figure}
\be
C^{(2,0)}+C^{(2,1)} = \sum_{n=1}^\infty \sum_{k=0}^6 c_{n,k} \, L^k_s \, z^{-n}
\; , \qquad
L_s \equiv \ln\left( -\frac{s}{M^2} - i0^+ \right) \; ,
\ee
\begin{align}
\begin{autobreak}
\mathcal{C}^{(2,0)}+\mathcal{C}^{(2,1)} =
\Big(54.56087661
-105.9626626 \, L_s
+8.887259013 \, L_s^2
+6.645715659 \, L_s^3
+0.8289545430 \, L_s^4
+0.03333333333 \, L_s^5
-0.001851851852 \, L_s^6\Big) \, z^{-1}
+\Big(85.66611966
+12.14331252 \, L_s
-34.87431988 \, L_s^2
-11.29224411 \, L_s^3
-0.8617012615 \, L_s^4
+0.01718750000 \, L_s^5
+0.001099537037 \, L_s^6\Big) \, z^{-2}
+\Big(4.242650309
+54.17459631 \, L_s
+4.950031592 \, L_s^2
+1.341075352 \, L_s^3
-0.9050441963 \, L_s^4
-0.02955729167 \, L_s^5
-0.001410590278 \, L_s^6\Big) \, z^{-3}
+\Big(-99.58994875
+39.36666234 \, L_s
+16.69670684 \, L_s^2
-2.994924396 \, L_s^3
-0.4321117491 \, L_s^4
+0.07813946759 \, L_s^5
-0.002267795139 \, L_s^6\Big) \, z^{-4}
+\Big(-13.01069569
-30.23605994 \, L_s
-33.09478959 \, L_s^2
+15.42454676 \, L_s^3
-0.1114137280 \, L_s^4
-0.08451605903 \, L_s^5
-0.004180230035 \, L_s^6\Big) \, z^{-5}
+\Big(-50.77508814
+315.6787714 \, L_s
-32.49111907 \, L_s^2
-27.01662191 \, L_s^3
+0.2417871470 \, L_s^4
+0.2111745877 \, L_s^5
+0.004659921152 \, L_s^6\Big) \, z^{-6}
+\Big(-220.1240019
-1889.592657 \, L_s
+543.4130382 \, L_s^2
+67.78095852 \, L_s^3
-9.594792355 \, L_s^4
-0.05716437042 \, L_s^5
-0.02886098226 \, L_s^6\Big) \, z^{-7}
+\Big(488.0690355
+11655.54167 \, L_s
-4220.431798 \, L_s^2
-99.26817518 \, L_s^3
+67.30519611 \, L_s^4
-2.224095015 \, L_s^5
+0.08744108412 \, L_s^6\Big) \, z^{-8}

+ \order{z^{-9}} \; .
\end{autobreak}
\end{align}
The parts of the numerical coefficients of terms proportional to $L_s^k/z$ for $k\in \lbrace 6,5,4,3\rbrace$, which stem from $\mathcal{C}^{(2,0)}$, comply with the exact coefficients predicted recently in Refs.~\cite{Liu:2020tzd, Liu:2020wbn}.


\section{Supplemental material} \label{sec:supplemental}

The supplemental material, Ref.~\cite{mathfile}, in form of a single file can be imported in \textsc{Wolfram Mathematica} for subsequent analysis. All variables are explained in the header of this file. The main function returns the form factor as a series in \verb|aspi|$\equiv \alpha_s/\pi$:
\begin{description}
\item \verb|CHaa[z, nh, nl, QQsum, Lmu]| - $\mathcal{C}$, Eq.~\eqref{eq:expansion};
\end{description}
The function \verb|CHaa[z, nh, nl, QQsum, Lmu]| is entirely based on the aforementioned expansions and interpolations. Hence, the analytical results of Ref.~\cite{Harlander:2019ioe} are not exploited. For the benefit of the reader, we provide all constituents of the form factor in terms of expansions and interpolation tables. The following functions are evaluated at $L_\mu =0$:
\begin{description}
\item \verb|C0[z], C1[z], C2[z, nh, nl, QQsum]| - $\mathcal{C}^{(0)}$, $\mathcal{C}^{(1)}$ and $\mathcal{C}^{(2)}$, Eqs.~\eqref{eq:expansion}~and~\eqref{eq:expansionC2};
\item \verb|C2LMEn00[z], C2LMEnh1[z], C2LMEnl1[z], C2LMEQQsum[z]| \\
- large-mass expansion of $\mathcal{C}^{(2,0)}$, $\mathcal{C}^{(2,1)}$ (Appendix~\ref{sec:LME}), $\mathcal{C}^{(2,2)}$ and $\mathcal{C}^{(2,3)}$;
\item \verb|C2THRn00[z], C2THRnh1[z], C2THRnl1[z], C2THRQQsum[z]| \\
- threshold expansion of $\mathcal{C}^{(2,0)}$, $\mathcal{C}^{(2,1)}$ (Appendix~\ref{sec:THR}), $\mathcal{C}^{(2,2)}$ and $\mathcal{C}^{(2,3)}$;
\item \verb|C2HEn00[z], C2HEnh1[z], C2HEnl1[z], C2HEQQsum[z]| \\
- high-energy expansion of $\mathcal{C}^{(2,0)}$, $\mathcal{C}^{(2,1)}$ (Appendix~\ref{sec:HE}), $\mathcal{C}^{(2,2)}$ and $\mathcal{C}^{(2,3)}$;
\item \verb|C2TABn00[z], C2TABnh1[z], C2TABnl1[z], C2TABQQsum[z]| \\
- interpolation of $\mathcal{C}^{(2,0)}$, $\mathcal{C}^{(2,1)}$ (Tabs.~\ref{tab:num1}~and~\ref{tab:num2}), $\mathcal{C}^{(2,2)}$ and $\mathcal{C}^{(2,3)}$.
\end{description}
One must supply numerical values for $z$. The large-mass expansion of $\mathcal{C}^{(2)}\big[ L_\mu = 0 \big]$ with exact coefficients can be called with \verb|C2LME|. Its dependence on $n_h$, $n_l$ and the sum over electric charges is kept variable.

\noindent
All functions are evaluated with fixed gauge group constants $C_A = 3$, $C_F = 4/3$, $T_F = 1/2$.


\newpage

\bibliographystyle{JHEP}
\bibliography{Haa}

\end{document}